\documentclass{article}

\newcommand{\tr}{\mbox{tr}}
\newcommand{\bra}[1]{\langle #1\vert}
\newcommand{\ket}[1]{\vert #1\rangle}
\newcommand{\braket}[2]{\langle #1\vert #2\rangle}
\begin{document}

\title{Classical Extensions, Classical Representations and Bayesian Updating in Quantum Mechanics}

\author{Guido Bacciagaluppi\\
IGPP, Wilhelmstraße 3a\\
D-79098 Freiburg i.Br., Germany\\
E-mail: gb@igpp.de}
\date{}

\maketitle

\begin{abstract}
I review the formalism of classical extensions of quantum mechanics introduced by Beltrametti and Bugajski,
and compare it to the classical representations discussed e.g.\ by Busch, Hellwig and Stulpe and recently used
by Fuchs in his discussion of quantum mechanics in terms of standard quantum measurements. I treat the problem of 
finding Bayesian analogues of the state transition associated with measurement in the canonical classical 
extension as well as in the related `uniform' classical representation. In the classical extension, the analogy 
is extremely good.   
\end{abstract}

\section{Introduction}\label{introduction}
One of the questions that have been especially discussed in this conference 
is that of the various ways of viewing the relation between 
classical probability and probability as it arises in quantum 
mechanics. The classical extensions of quantum mechanics 
studied by Beltrametti and Bugajski\nolinebreak\cite{BelBug1995,Bug1996,BelBug1996,BelBug2002,BelBug2003} 
are an example of a formalism providing a well-defined way 
of seeing quantum probabilities as a special case of 
classical probabilities. The first aim of this paper is to give a 
short review of this formalism and compare it to another approach 
that could be described roughly as doing the same, 
the so-called classical representations of quantum mechanics,\nolinebreak\cite{BusHelStu1993} 
which are used by Chris Fuchs in his analysis of quantum mechanics in terms of 
standard quantum measurements.\nolinebreak\cite{Fuc2002} 
The second aim is to examine further the possible relation between 
quantum measurement and Bayesian updating, also a focus of
Fuchs's discussion (which in fact provided the original impetus for 
this talk).

Paul Busch's paper given at this conference\nolinebreak\cite{Bus2003} 
also deals with the formalism of classical extensions, so that, 
in Section \ref{formalism}, I can limit myself to sketching the 
general lines of the approach. In Section \ref{simplex}, I look 
at classical extensions as compared to classical representations, 
focusing on the question of Bayesian updating in Section 
\ref{updating}. Brief caveats about `classicality' (Section \ref{conclusion}) conclude the paper.
I deal throughout with quantum mechanics in {\em finite} dimensions.

As compared to the talk (simply entitled `Beltrametti and Bugajski's classical extensions
of quantum mechanics') this paper includes a fuller discussion in Sections \ref{simplex} and \ref{updating}, 
in particular new positive results.

\section{Formalism}\label{formalism}
The convex set approach aims at representing the operational aspects of
physical theories. The primary notion 
is that of a convex set of states $S$, e.g.\ the density operators, 
or the probability measures on some phase space $\Omega$. An observable 
is defined to be a suitable affine mapping from $S$ to the space 
$M_1^+(\Xi)$ of probability measures on a corresponding value space $\Xi$. 
Equivalently, one can also define an observable as an effect-valued 
measure on a value space $\Xi$, where an effect is a suitable affine 
map from $S$ to the interval $[0,1]$. The sets of states and observables 
are chosen such that states separate observables and 
observables separate states, in the usual sense that there are enough 
states to distinguish different observables and vice versa.   

As a familiar example, one can take the case of standard quantum 
mechanics. Here $S$ is the set of density operators $\rho$ on some
Hilbert space, and an effect $E$ is the map from $S$ to $[0,1]$ given by
  \begin{equation}
    \rho\mapsto\tr(\rho E),
    \label{guido1}
  \end{equation}
where $E$ is a positive operator with spectrum in the interval $[0,1]$ (also 
called {\em effect}). An effect-valued measure (observable), also
called positive-operator-valued measure (POV measure or POVM), is an 
association of sets $X$ from a $\sigma$-algebra of subsets of $\Xi$ with
effects $E_X$, with suitable properties of normalisation and 
$\sigma$-additivity, such that applied to a state $\rho$ it will induce
an ordinary probability measure on $\Xi$:
  \begin{equation}
    X\mapsto p(X)=\tr(\rho E_X).
    \label{guido2}
  \end{equation} 
As a special case of observables, one can consider measures where 
all $E_X$ are in fact projections (projection-valued or 
PV measure, or PVM). As is well known, this corresponds to the definition 
of an observable as a self-adjoint operator (which defines a unique PV
measure through the spectral theorem).

\subsection{The classical case}
The application of this formalism to a classical setting results in what
Bugajski calls fuzzy probability theory.\nolinebreak\cite{Bug1996} This may be 
less familiar than the quantum example, but will be of particular interest in the following.
In a classical setting, the convex set $S$ is a simplex, specifically 
the space $M_1^+(\Omega)$ of probability measures over
some phase space $\Omega$. (For the purposes of the following, we 
can restrict attention to `regular' or `measurable' observables
and effects.)

A (regular) effect on $S$ can be identified with a function $e(\omega)$ 
from $\Omega$ to $[0,1]$ in the sense that it defines an affine mapping
from $S$ to $[0,1]$ via
  \begin{equation}
    \mu\mapsto\int e(\omega)d\mu(\omega).
    \label{guido3}
  \end{equation}
An observable is then an effect-valued measure $X\mapsto e_X(\omega)$ with 
the usual properties, such that when applied to a
state $\mu$ it induces an ordinary (normalised) probability measure:
  \begin{equation}
    X\mapsto p(X)=\int_\Omega e_X(\omega)d\mu(\omega).
    \label{guido4}
  \end{equation}
A special case are the observables obtained when the effects $e_X$ 
are characteristic functions. As a matter of fact, this corresponds 
to the more standard definition of an observable as a random variable 
$f:\Omega\rightarrow\Xi$, via the correspondence
  \begin{equation}
    e_X(\omega)=\chi_{f^{-1}(X)}(\omega).
    \label{guido5}
  \end{equation}

Thus, general observables as defined in the convex set approach 
in the classical case will be {\em fuzzy random variables}, which are
not dispersion-free for all pure states (in operational parlance, 
they produce indeterministic results upon measurement even in certain 
pure states). Indeed, the probability measure obtained by 
application to the pure state $\delta_{\omega_0}$ (with fixed $\omega_0$)
is $p(X)=e_X(\omega_0)$, so that pure 
states (delta measures) are generally not mapped into pure states
(delta measures). While this feature appears also in the quantum 
mechanical case (in fact there for {\em all} observables),
one can show that for any two classical (regular) observables 
there exists (non-uniquely) a {\em joint} observable, due to the fact 
that one can always construct probability measures with given marginals.

\subsection{Extensions of state spaces}\label{extension}
Take two state spaces $S_1$ and $S_2$. By definition, $S_1$ will 
extend $S_2$ iff there is an affine map 
  \begin{equation}
    R:\quad S_1\rightarrow S_2
    \label{guido6}
  \end{equation}
that is {\em surjective} (called the reduction map). Observables on $S_2$ will 
correspondingly induce observables on $S_1$:
  \begin{equation}
    A:\quad S_2\rightarrow M_1^+(\Xi)\quad\mbox{(affine)}
    \label{guido7} 
  \end{equation}
induces
  \begin{equation}
    A\circ R:\quad S_1\rightarrow M_1^+(\Xi)\quad\mbox{(affine)},
    \label{guido8}
  \end{equation}
and the two observables define exactly the same statistics on 
corresponding states, e.g.\ $A$ in the state $R(s)$ will have the same
dispersion as $A\circ R$ in the state $s$, or a joint observable 
of $A$ and $B$ (if it exists) will map to a joint observable of $A\circ R$
and $B\circ R$, etc.

A familiar example of such an extension of the state space is obtained taking 
$S_1$ as the density operators for some quantum system,
$S_2$ the density operators for a subsystem and $R$ the partial 
trace. A few results that are easy to establish in general (and 
are evident in the example) are:
  \begin{itemize}
    \item mixed states map to mixed states;
    \item pure states can map to pure or to mixed states;
    \item pure states have unique preimages (which are pure), and
    \item mixed states can have several preimages.
  \end{itemize}
Intuitively, an extension $S_1$ will have more states than $S_2$, 
and also more observables than those induced by $S_2$-observables,
since there must be enough $S_1$-observables to separate the states.

\subsection{The canonical classical extension of quantum mechanics}\label{canonical}
Beltrametti and Bugajski,\nolinebreak\cite{BelBug1995} extending work by 
Misra,\nolinebreak\cite{Mis1974} discuss a classical extension of quantum 
mechanics, defined as follows. Let $S$ be the convex set of 
density operators on some Hilbert space, and let $M_{1}^{+}(\partial S)$ be 
the set of classical probability measures on the set of extremal 
points of $S$ (the set of pure states of the quantum system). Define 
the reduction map
  \begin{equation}
    R:\quad M_{1}^{+}(\partial S)\rightarrow S
    \label{guido9}
  \end{equation}
as follows: map the pure states bijectively to the corresponding 
pure states, and extend by affinity, i.e.\ map convex combinations
to the corresponding convex combinations.

Explicitly (for a finite-dimensional quantum system), 
this amounts to:
  \begin{equation}
     \delta_{\omega'}(\omega)\mapsto
     \int\delta_{\omega'}(\omega)\ket{\omega}\bra{\omega}d\omega
                                            =\ket{\omega'}\bra{\omega'}
    \label{guido10}
  \end{equation}
(one-one correspondence), and
  \begin{equation}
     p(\omega)\mapsto\int p(\omega)\ket{\omega}\bra{\omega}d\omega
    \label{guido11}
  \end{equation}
(many-one), where $d\omega$ is the (normalised) unitarily invariant measure 
on the hypersphere (e.g.\ the Bloch sphere). Notice that classical states that map to the same 
mixed quantum state correspond to that state's different convex 
decompositions into pure quantum states. (Henceforth we shall 
often write q-states and c-states for quantum and classical states, 
respectively; similarly for q-observables and c-observables, etc.)

Observables (i.e.\ POVMs) on the quantum states $S$ now will 
induce observables on the classical states $M_1^+(\partial S)$ with 
the same statistics. Notice that these are not {\em all} the 
c-observables, since c-observables separate c-states
and could thus be used to distinguish different convex decompositions of a
q-state. Explicitly, the induced effects are as follows. Let $E$ be an effect 
on the q-states, i.e. the affine mapping into $[0,1]$ given by
  \begin{equation}
    \rho\mapsto\tr(\rho E),
    \label{guido12}
  \end{equation}
and let $p(\omega)$ be any c-state reducing to $\rho$ under $R$,
i.e.\ $\rho=R[p(\omega)]$. Inserting (\ref{guido11}) for $\rho$ in (\ref{guido12}) yields
  \begin{equation}
      \int p(\omega)\ket{\omega}\bra{\omega}d\omega 
      \mapsto
      \tr\left(\int p(\omega)E\ket{\omega}\bra{\omega}d\omega
         \right)=
      \int p(\omega)\bra{\omega}E\ket{\omega}d\omega. 
    \label{guido13}
  \end{equation}
From this we can see that the q-effect $E$ induces a corresponding c-effect $E\circ R$ given by 
  \begin{equation}
    e(\omega)= \bra{\omega}E\ket{\omega}, 
    \label{guido14}
  \end{equation}
or more precisely by the affine map from the c-states to $[0,1]$ defined by 
  \begin{equation}
    p(\omega)\mapsto\int e(\omega)p(\omega)d\omega,
    \label{guido15}
  \end{equation}
with $e(\omega)$ as in (\ref{guido14}). 

For all pairs of corresponding states $p(\omega)$ and $\rho=R[p(\omega)]$ and
pairs of corresponding effects $E$ and $e(\omega)=E\circ R$ we have
  \begin{equation}
    \tr(\rho E)=\int e(\omega)p(\omega)d\omega,
    \label{guido16}
  \end{equation}
that is, they share the same statistics. The corresponding result for
observables (effect-valued measures) follows from (\ref{guido16}). In 
particular, every q-observable has a corresponding c-observable 
with the same statistics in the classical extension, which we shall call 
the c-representative of the q-observable.

Below are listed a few immediate results. While at first they may seem surprising, 
the key to understanding them is the fact that c-representatives of 
q-observables are {\em always} fuzzy;
indeed, the c-effect $ e(\omega)= \bra{\omega}E\ket{\omega}$ is
never a characteristic function, not even if $E$ is a projection:
  \begin{itemize}
    \item c-representatives of incompatible q-observables obey the same
          dispersion (uncertainty) relations;
    \item c-representatives of any two q-observables have a joint
          c-observable (this is not the c-representative of any
          q-observable if the two q-observables do not have a joint
          q-observable);
    \item c-representatives of certain single and joint q-observables
          violate the Bell inequalities.
  \end{itemize}
These and related points are further elaborated by Beltrametti and 
Bugajski, but we cannot review all of them here. In particular,\nolinebreak\cite{BelBug1996} 
they discuss violation of the Bell inequalities, or what they call in general the 
Bell phenomenon in fuzzy probability theory (see also Section \ref{conclusion}).  
They also discuss further aspects of what they call probabilistic entanglement, in particular 
the distinction between classical and quantum correlations.\nolinebreak\cite{BelBug2002,BelBug2003}

The c-extension of quantum mechanics appears to provide us with a new formulation
of quantum mechanics, one in which the states are the convex decompositions 
of the standard q-states, but the c-observables on these 
states are suitably restricted, thus ensuring that different convex
decompositions of a q-state are operationally indistinguishable
(so to speak a `hidden observables' theory). 
Indeed, quantum mechanics seems to reduce to a special case of
fuzzy probability theory, at least as far as statistical predictions
are concerned. The further issue of the updating
of the c-state upon measurement, however, needs to be clarified, and we shall
discuss it in Section \ref{updating}.

\section{Comparison with classical representations}\label{simplex}
Another formalism that represents quantum mechanics in terms 
of (restrictions on) classical probabilities
has recently enjoyed the limelight in the context of the 
possible information-theoretic foundation of
quantum mechanics and of the connections between quantum 
measurement and Bayesian updating, especially in work by 
Fuchs.\nolinebreak\cite{Fuc2002} The formal framework, 
previously discussed under the heading of classical representations 
of quantum mechanics,\nolinebreak\cite{BusHelStu1993} will be briefly 
summarised below, and relies on describing quantum states 
as defining distributions over the values of certain POVMs. 
While such a framework can be easily construed as a beable theory
(a theory describing  distributions over actually existing 
values of the POVM), Fuchs's working hypothesis is that it 
should be construed instead as a representation of information 
in terms of results of some `standard quantum measurement'. Since 
no quantum state provides complete information about these results, 
this prompts the question of finding (information-theoretic) 
reasons that might constrain the probability measures to be 
the special measures representing quantum states.

It is natural to expect that also the formalism of classical 
extensions could be used in this kind of programme. Starting with a state
space $\Omega$, not yet endowed with a Hilbert-space structure, 
with c-states $M_1^+(\Omega)$ and with the corresponding 
c-observables, one would need to find constraints on the 
{\em observables} rather than the states, until one arrives to the 
c-extension of a quantum system.  In this sense, the formalism 
of classical extensions may be interesting for the purpose of 
investigating which aspects of quantum mechanics might be 
reproduced or simulated using classical systems. The discussion
below should partly clarify the scope and limitations of such
a project (see also Section \ref{conclusion}).

Classical representations of quantum mechanics in $n$ 
dimensions\nolinebreak\cite{BusHelStu1993} are based on the fact that
for certain, say, discrete POVMs $X\mapsto\sum_{i\in X}E_i$, the effects $E_i$ form a basis
in the $n^2$-dimensional real vector space of self-adjoint operators.
The probability distribution on the value space of the POVM defined
by a density operator $\rho$, i.e.\ $\tr(\rho E_i)$, thus yields the
`projections' of the operator $\rho$ onto the `axes' $E_i$, and can be
used to uniquely reconstruct $\rho$.\footnote{That is, the measurement
statistics for the POVM are enough to reconstruct completely any
quantum state $\rho$, unlike the case of a PV measure, which defines
at most $n$ `axes'. For my take on the meaning of measurements of POVMs,
see the paper by Cattaneo {\em et al.},\nolinebreak\cite{Cat1997} Sections 4 and 5.}

The probabilities $\tr(\rho E_i)$ should be distinguished from the coefficients
$\rho_i$ in an {\em expansion} of $\rho$ in terms of the basis operators $E_i$:
  \begin{equation}
    \rho=\sum_{i=1}^N\rho_iE_i.
    \label{guido16a}
  \end{equation}
Indeed, one has the relation
  \begin{equation}
    \tr(\rho E_j)=\tr\left(\sum_{i=1}^N\rho_iE_iE_j\right)=\sum_{i=1}^N\rho_i\tr(E_iE_j),
    \label{guido16b}
  \end{equation}
i.e.\ the probability distribution $\tr(\rho E_i)$ is a smearing out of the function
$\rho_i$. Specialising to the case where the $E_i$ have the form
  \begin{equation}
    E_i=\frac{1}{\Omega}\ket{\omega_i}\bra{\omega_i}
    \label{guido16c}
  \end{equation}
for some normalised vectors $\ket{\omega_i}$, with $\Omega$ determined by the 
normalisation condition $\sum_{i=1}^NE_i={\bf 1}$, i.e.
  \begin{equation}
    \tr\left(\sum_{i=1}^NE_i\right)=\sum_{i=1}^N\frac{1}{\Omega}=\frac{N}{\Omega}=\tr({\bf 1})=n,\\[0.5ex]
    \label{guido16d}
  \end{equation}
we have
  \begin{equation}
    \frac{1}{\Omega}\tr(\rho\ket{\omega_j}\bra{\omega_j})=
    \sum_{i=1}^N\rho_i\frac{1}{\Omega^2}\tr(\ket{\omega_i}\bra{\omega_i}\ket{\omega_j}\bra{\omega_j}),
    \label{guido16e}
  \end{equation}
or
  \begin{equation}
    \frac{1}{\Omega}\bra{\omega_j}\rho\ket{\omega_j}=
    \sum_{i=1}^N\rho_i\frac{1}{\Omega^2}|\braket{\omega_i}{\omega_j}|^2,
    \label{guido16f}
  \end{equation}
and we see that this must always be a true smearing out, since 
$\frac{1}{\Omega^2}|\braket{\omega_i}{\omega_j}|^2=\delta_{ij}$ would imply that there are $N$
($\geq n^2$) orthogonal vectors in an $n$-dimensional Hilbert space.

Also, notice that while $\tr(\rho E_i)$ is uniquely determined, the coefficients
$\rho_i$ will be unique only if $N=n^2$, i.e. if the basis elements $E_i$ are linearly independent
(form a {\em minimal} informationally complete POVM). Otherwise, the basis is overcomplete,
and the representation (\ref{guido16a}) is non-unique.

Finally, and most importantly, the function $\rho_i$ appearing in (\ref{guido16a}) is in
general only a pseudo probability distribution, i.e. it can take negative 
values.\nolinebreak\cite{BusHelStu1993} This can be made intuitively clear if we again specialise
to the case of (\ref{guido16c}). In this case, the affine mapping, call it $P$, from
the true probability measures on $\{1,\ldots,N\}$ into the density operators, defined
by
  \begin{equation}
    P:\quad p\mapsto\sum_ip_i\ket{\omega_i}\bra{\omega_i}=\Omega\sum_i p_i E_i,
    \label{guido16g}
  \end{equation}
is not surjective, because the only pure states in the image of $P$ are the states
$\ket{\omega_i}\bra{\omega_i}$ themselves. Therefore, for some quantum states the
representation (\ref{guido16a}) is {\em not} given by coefficients of the form
$p_i$, with $p_i$ a probability distribution, nor indeed with $p_i\geq 0$
for all $i$, and must before become negative.\footnote{The relation between a non-positive 
$\rho_i$ and the probability distribution $\frac{1}{\omega}\bra{\omega_i}\rho\ket{\omega_i}$ 
is quite analogous to that between the non-positive Wigner function and true phase space 
probability distributions such as the Husimi function, which are defined using POVMs 
of coherent states (see Section VI.1 of the textbook by Busch, Grabowski and Lahti).\nolinebreak\cite{BusGraLah1995}} 

While $P$ is in some ways analogous to the reduction map $R$ of Section \ref{extension},
the recourse to pseudo probability distributions prompts one to talk of a {\em pseudo
classical extension} as opposed to the classical extension of Section \ref{canonical},
and as opposed to the {\em classical representation} in terms of the
true probability distribution $\tr(\rho E_i)=\frac{1}{\Omega}\bra{\omega_i}\rho\ket{\omega_i}$.

In the analysis of the quantum state as information about a standard quantum measurement 
(Section 4 of his paper\nolinebreak\cite{Fuc2002}), Fuchs suggests to take as the informationally complete
reference POVM one of the form (\ref{guido16c}) that is both symmetric (in terms of the
scalar products of the vectors $\ket{\omega_i}$) and minimal ($N=n^2$).

For the purpose of comparing Fuchs's analysis with Beltrametti and Bugajski's c-extension, 
it will be more expedient to use a `maximal' POVM, whose 
value space is the set of all pure states of the quantum system under 
consideration, namely the so-called uniform POVM: 
  \begin{equation}
    X\mapsto\int_X\ket{\omega}\bra{\omega}d\omega,
    \label{guido27}
  \end{equation}
where $d\omega$ is again the normalised unitarily invariant measure 
on the hypersphere. Quantum states are 
thus represented as a {\em subset} of the probability measures $M_{1}^{+}(\partial S)$, i.e.\ 
a subset of the measures used in the c-extension of quantum mechanics.

Explicitly, a pure state $\ket{\psi}$ will have its classical 
representation given by the density
  \begin{equation}
    \psi(\omega)=|\braket{\psi}{\omega}|^2,
    \label{guido28}
  \end{equation}
and a mixed state $\rho$ by
  \begin{equation}
    \rho(\omega)=\bra{\omega}\rho\ket{\omega}.
    \label{guido29}
  \end{equation}

Now, however, the probability measures over the value space 
of the uniform POVM are mapped surjectively onto the quantum states
$\rho$ by the reduction map $R$, rather than non-surjectively by 
the corresponding map $P$ in the case of a minimal POVM. Thus the
coefficients in the expansions in terms of the uniform POVM (which 
for mixed states are vastly non-unique) are indeed true probability 
distributions, the distributions defined in the c-extension of 
quantum mechanics. Taking any $p(\omega)$ such that
  \begin{equation}
    \rho=\int p(\omega)\ket{\omega}\bra{\omega}d\omega,
    \label{guido30}
  \end{equation}
the smearing out relation between $p(\omega)$ and the classical 
representation of $\rho$ now takes the form
  \begin{equation}
    \bra{\tilde{\omega}}\rho\ket{\tilde{\omega}}=
    \int p(\omega)|\braket{\omega}{\tilde{\omega}}|^2d\omega,
    \label{guido31}
  \end{equation}
i.e.\ a smearing out of $p(\omega)$ with the function
  \begin{equation}
    S(\omega,\tilde{\omega})=|\braket{\omega}{\tilde{\omega}}|^2.
    \label{guido32}
  \end{equation}
When the q-states are represented using the uniform classical 
representation, this gives us an explicit form of the reduction map 
$p(\omega)\mapsto\rho(\omega)$, from the c-states of the canonical 
extension to the q-states, as a 
smearing out of $p(\omega)$, indeed illustrating the fact that
probability distributions in the classical representation are never pure.

\section{Quantum measurement and Bayesian updating}\label{updating}
One of the topics under discussion in Fuchs's paper\nolinebreak\cite{Fuc2002} (his Section 6) 
is the possible relation between quantum measurement (in the sense of the 
transformation of the state upon measurement, however caused, usually 
called the `collapse' of the quantum state) and
Bayesian updating of probability distributions. 

Collapse in the case of measurement of some (discrete) POVM 
using (pure) operations $A_d$ (with $A_d^*A_d=E_d$) takes the form
  \begin{equation}
    \rho\mapsto\rho^d=\frac{A_d\rho A_d^*}{\tr(A_d\rho A_d^*)}
    \label{guido18}
  \end{equation}
with probability
  \begin{equation}
    \quad\tr(\rho E_d)=\tr(A_d\rho A_d^*).
    \label{guido18a}
  \end{equation}
A disanalogy with Bayes' rule lies in the fact that in general
  \begin{equation}
    \rho\neq\sum_d\tr(\rho E_d)\rho^d,
    \label{guido19}
  \end{equation}
so there appears to be no direct interpretation of collapse as a
{\em selection} of a term in a convex decomposition of the initial state. 
On the other hand, Fuchs points out that the operator 
$\sum_d A_d\rho A_d^*$ is unitarily equivalent to
  \begin{equation}
      \sum_d\rho^{1/2}A_d^*A_d\rho^{1/2}=\sum_d\rho^{1/2}E_d\rho^{1/2}=\rho,
    \label{guido20}
  \end{equation}
so that one can instead reinterpret `collapse' as a selection of a term
  \begin{equation}
    \tilde{\rho}^d=\rho^{1/2}A_d^*A_d\rho^{1/2}
    \label{guido21}
  \end{equation}
in the convex decomposition (\ref{guido20}) of $\rho$, followed by
a unitary `readjustment'
  \begin{equation}
    \tilde{\rho}^d\mapsto\rho^d=V_d\tilde{\rho}^dV_d^*,
    \label{guido22}
  \end{equation}
for suitable $V_d$ (which in general depends on {\em both} $\rho$ and $E_d$). 
Thus it becomes possible to see collapse as a non-commutative variant of Bayes' 
rule.\footnote{Notice that one can also represent the collapse as a selection of
a term in a decomposition of the state of the system {\em after} the appropriate
interaction with a measuring apparatus.}

The corresponding transformation in the classical representation 
is obtained if we substitute the probability 
distribution $\frac{1}{\Omega}\bra{\omega_i}\rho\ket{\omega_i}$ for $\rho$ in the above. 
The standard collapse becomes
  \begin{equation}
    \frac{1}{\Omega}\bra{\omega_i}\rho\ket{\omega_i}\mapsto \frac{1}{\Omega}\bra{\omega_i}\rho^d\ket{\omega_i}=
    \frac{1}{\Omega}\frac{\bra{\omega_i}A_d\rho A_d^*\ket{\omega_i}}{\tr(\rho E_d)}
    \label{guido23}
  \end{equation}
with probability $\tr(\rho E_d)$. Or in the reinterpretation,
  \begin{equation}
    \frac{1}{\Omega}\bra{\omega_i}\rho\ket{\omega_i}\mapsto \frac{1}{\Omega}\bra{\omega_i}\tilde{\rho}^d\ket{\omega_i}=
    \frac{1}{\Omega}\frac{\bra{\omega_i}\rho^{1/2}E_d\rho^{1/2}\ket{\omega_i}}{\tr(\rho E_d)}
    \label{guido24}
  \end{equation}
with probability $\tr(\rho E_d)$, followed by the unitary readjustment, where now indeed
  \begin{equation}
    \frac{1}{\Omega}\bra{\omega_i}\rho\ket{\omega_i}=
    \sum_d\tr(\rho E_d)\frac{1}{\Omega}\bra{\omega_i}\tilde{\rho}^d\ket{\omega_i}.
    \label{guido25}
  \end{equation}
In a sense thus we have an application of the standard Bayes rule, followed by the readjustment 
$\frac{1}{\Omega}\bra{\omega_i}\tilde{\rho}^d\ket{\omega_i}\mapsto\frac{1}{\Omega}\bra{\omega_i}\rho^d\ket{\omega_i}$, 
which can also be interpreted passively as a unitary readjustment of 
the reference POVM:
  \begin{equation}
    \frac{1}{\Omega}\ket{\omega_i}\bra{\omega_i}\mapsto 
    \frac{1}{\Omega}\ket{\omega_i^d}\bra{\omega_i^d}:=
    V_d^*\frac{1}{\Omega}\ket{\omega_i}\bra{\omega_i}V_d.
    \label{guido26}
  \end{equation}

The analogy, however, as duly emphasised by Fuchs, does not extend to the 
reinterpretation of the selected component as
  \begin{equation}
    \frac{1}{\Omega}\bra{\omega_i}\tilde{\rho}^d\ket{\omega_i}=
    \frac{e_d^i\frac{1}{\Omega}\bra{\omega_i}\rho\ket{\omega_i}}{\sum_ie_d^i\frac{1}{\Omega}\bra{\omega_i}\rho\ket{\omega_i}},
    \label{guido32b}
  \end{equation}
with $e_d^i$ being the characteristic function of some set. Indeed, the transition
resists such a reinterpretation even if we allow $e_d^i$ to be a (fuzzy) classical
effect.

One reason for this is surely that, while in a minimal classical
representation, for any given effect $E_d$ there is a {\em unique} function 
$e_d^i$ such that for all $\frac{1}{\Omega}\bra{\omega_i}\rho\ket{\omega_i}$,
  \begin{equation}
    \tr(\rho E_d)=\sum_{i=1}^Ne_d^i\frac{1}{\Omega}\bra{\omega_i}\rho\ket{\omega_i},
    \label{guido33}
  \end{equation}
this function $e_d^i$ will in general not be positive, and thus not a classical
effect.\nolinebreak\cite{BusHelStu1993} Indeed, since $E_d$ is self-adjoint, one can
represent it as
  \begin{equation}
    E_d=\frac{1}{\Omega}\sum_{i=1}^Ne_d^i\ket{\omega_i}\bra{\omega_i}
    \label{guido34}
  \end{equation}
with suitable $e_d^i$ (unique if $N=n^2$). Also,
  \begin{equation}
    \tr(\rho E_d)=\tr\left(\rho \frac{1}{\Omega}\sum_{i=1}^Ne_d^i\ket{\omega_i}\bra{\omega_i}\right)
                =\sum_{i=1}^Ne_d^i\frac{1}{\Omega}\bra{\omega_i}\rho\ket{\omega_i}.
    \label{guido35}
  \end{equation}
However, since $E_d$ is proportional to a density operator, and in general the representation
(\ref{guido16a}) of a density operator is in terms of a non-positive function, so is the
representation (\ref{guido34}), and $e_d^i$ in general is not a classical effect.

The same analysis can be repeated using the uniform POVM, and just as one has true 
probability distributions intead of pseudo probability distributions in the classical
extension, so one has classical effects appearing in the expansions of the form
  \begin{equation}
    E_d=\int e_d(\omega)\ket{\omega}\bra{\omega}d\omega.
    \label{guido36}
  \end{equation}
In fact, it is easy to see that $\tr(\rho E_d)$ is interpretable equally well as
  \begin{equation}
    \tr(\rho E_d)=\tr\left(\rho\int e_d(\omega)\ket{\omega}\bra{\omega}d\omega\right)
                =\int e_d(\omega)\bra{\omega}\rho\ket{\omega}d\omega
    \label{guido37}
  \end{equation}
(classical representation), or as
  \begin{equation}
    \tr(E_d\rho)=\tr\left(E_d\int p(\omega)\ket{\omega}\bra{\omega}d\omega\right)
                =\int p(\omega)\bra{\omega}E_d\ket{\omega}d\omega
    \label{guido38}
  \end{equation}
(classical extension), where, respectively, $e_d(\omega)$ and $\bra{\omega}E_d\ket{\omega}$
are classical effects, the latter obtained from the former again by {\em smearing out} with the 
function (\ref{guido32}).

Thus one might hope that {\em in the classical extension}, or in the related {\em classical
representation} that uses the uniform POVM, the transition $\rho\mapsto\tilde{\rho}^d$
might indeed be interpretable in terms of a {\em fully classical} Bayesian updating,
at least under an appropriate choice of the (non-unique) $p(\omega)$ reducing to $\rho$ 
in the c-extension, or of the (non-unique) $e_d(\omega)$ in the expansion of $E_d$
in the c-representation.\footnote{We record for posterity the phrase `essential onticity', which
Lucien Hardy coined in this connection.} A further bonus would be that in the case 
of the uniform POVM the subsequent unitary readjustment $\tilde{\rho}^d\mapsto\rho^d$ could be 
interpreted in the passive version as a relabelling of the vectors in the uniform POVM, rather
than a change of reference POVM altogether.

By means of examples, one can easily see that these hopes are misguided. One can, however, 
provide the following analyses in the c-representation and c-extension, respectively, which
arguably come close to simple Bayesian updating. 

Take {\em any} representation of $E_d$ in terms of a classical effect, i.e.
  \begin{equation}
    E_d=\int e_d(\omega)\ket{\omega}\bra{\omega}d\omega.
    \label{guido39}
  \end{equation}
The transition $\rho\mapsto\rho^{1/2}E_d\rho^{1/2}$ can be written as
  \begin{equation}
    \begin{array}{rcl}
      {\displaystyle \rho\mapsto\rho^{1/2}\int e_d(\omega)\ket{\omega}\bra{\omega}d\omega\rho^{1/2} }      & = &
      {\displaystyle \int e_d(\omega)\rho^{1/2}\ket{\omega}\bra{\omega}\rho^{1/2}d\omega            }  \\  & = &
      {\displaystyle \int e_d(\omega)\rho(\omega)\ket{\sigma(\omega)}\bra{\sigma(\omega)}d\omega,   }
      \label{guido40}
    \end{array}
  \end{equation}
where
  \begin{equation}
    \ket{\sigma(\omega)}:=\left\{\begin{array}{ll}
                                   \frac{\rho^{1/2}\ket{\omega}}{\sqrt{\bra{\omega}\rho^{1/2}\rho^{1/2}\ket{\omega}}}=
                                   \frac{\rho^{1/2}\ket{\omega}}{\sqrt{\rho(\omega)}}  &  
                                                         \quad\mbox{if $\rho^{1/2}\ket{\omega}\neq 0$,}  \\[1.5ex]
                                   0     &    \quad\mbox{otherwise.}
                                 \end{array}
                          \right.
    \label{guido41}
  \end{equation}
The c-representation $\rho^{1/2}E_d\rho^{1/2}(\tilde{\omega})$ of this operator is
  \begin{equation}
    \bra{\tilde{\omega}}\left(\int e_d(\omega)\rho(\omega)\ket{\sigma(\omega)}\bra{\sigma(\omega)}d\omega\right)\ket{\tilde{\omega}}=
    \int e_d(\omega)\rho(\omega)|\braket{\tilde{\omega}}{\sigma(\omega)}|^2d\omega.
    \label{guido42}
  \end{equation}
That is, the transition $\rho\mapsto\tilde{\rho}^d(\omega)$ can be interpreted as a {\em Bayesian updating}
  \begin{equation}
    \rho(\omega)\mapsto\frac{e_d(\omega)\rho(\omega)}{\int e_d(\omega)\rho(\omega)d\omega}
    \label{guido43}
  \end{equation}
with a classical effect $e_d(\omega)$ depending only on $E_d$, followed by a {\em smearing out} with a function
$|\braket{\tilde{\omega}}{\sigma(\omega)}|^2$ depending only on $\rho$. The full transition 
$\rho(\omega)\mapsto\rho^d(\omega)$ further includes Fuchs's unitary readjustment.

In this sense, the analogy between quantum measurement in the {\em maximal} c-representation and the classical
Bayes rule goes further than in the case of a minimal c-representation (where the above analysis can also
be carried out, but as we have seen $e_d^i$ is not generally a positive function.

The analogy is even more pleasing in the case of the canonical c-extension. For this, we can consider directly 
the transition
  \begin{equation}
    \rho\mapsto\rho^d=\frac{1}{\tr(\rho E_d)}A_d\rho A_d^*.
    \label{guido44}
  \end{equation}
Take {\em any} classical probability distribution $p(\omega)$ that reduces to $\rho$:
  \begin{equation}
    \rho=\int p(\omega)\ket{\omega}\bra{\omega}d\omega.
    \label{guido45}
  \end{equation}
We can write
  \begin{equation}
    \begin{array}{rcl}
      A_d\rho A_d^*   & = &   {\displaystyle A_d\int p(\omega)\ket{\omega}\bra{\omega}d\omega A_d^*  }  \\
                      & = &   {\displaystyle \int p(\omega)A_d\ket{\omega}\bra{\omega}A_d^*d\omega   }  \\
                      & = &   {\displaystyle 
      \int p(\omega)\bra{\omega}E_d\ket{\omega}\ket{\alpha(\omega)}\bra{\alpha(\omega)}d\omega,  }
    \end{array}
    \label{guido46}
  \end{equation}
where, analogously to (\ref{guido41}),
  \begin{equation}
    \ket{\alpha(\omega)}:=\left\{\begin{array}{ll}
                                   \frac{A_d\ket{\omega}}{\sqrt{\bra{\omega}A_d^*A_d\ket{\omega}}}=
                                   \frac{A_d\ket{\omega}}{\sqrt{\bra{\omega}E_d\ket{\omega}}}  &  
                                                         \quad\mbox{if $A_d\ket{\omega}\neq 0$,}  \\[1.5ex]
                                   0     &    \quad\mbox{otherwise.}
                                 \end{array}
                          \right.
    \label{guido47}
  \end{equation}
That is, the transition $\rho\mapsto\rho^d$ can be interpreted as a {\em Bayesian updating}
  \begin{equation}
    p(\omega)\mapsto\frac{p(\omega)\bra{\omega}E_d\ket{\omega}}{\int p(\omega) \bra{\omega}E_d\ket{\omega}d\omega}
    \label{guido48}
  \end{equation}
depending only on $E_d$, followed by a non-linear and generally many-to-one {\em disturbance} of the $\ket{\omega}$
depending only on the operation $A_d$ and in particular not on $p(\omega)$ or $\rho$.

This seems to be a natural generalisation of Bayes' rule to the case in which the gathering of data causes a
disturbance of the system. With this addition to the classical updating procedure, the 
operational aspects of quantum mechanics can indeed be described fully within the framework of fuzzy 
probability theory.

\section{Conclusion}\label{conclusion}
Although the results by Beltrametti and Bugajski and the above may imply that the behaviour of single quantum
systems can be successfully simulated using (random) classical systems, it would be rash to jump to conclusions
about a return to a classical theory. As soon as one considers composite systems, one should expect departures
from classical intuition. Indeed, the phase space $\Omega$ corresponding to a composite system does not appear 
to have the structure of a Cartesian product $\Omega_1\times\Omega_2$, which violates classical intuitions about
separability. Also, the measurement statistics that reproduce the violations of the Bell inequalities in the 
canonical c-extension violate {\em outcome independence}, which violates classical intuitions about locality 
or classical intuitions about common causes. In this context, a comparison with Fuchs's results about Gleason's
theorem in composite systems\nolinebreak\cite{Fuc2002} (his Section 5) would seem useful, as well as a thorough comparison 
with the recent work by Spekkens.\nolinebreak\cite{Spe2003}

\section*{Acknowledgments}
I happily acknowledge my debt to Howard Barnum, Chris Fuchs, Lucien Hardy, Rob Spekkens and Alex Wilce
for exciting and useful discussions.

\end{document}